\documentclass[10pt]{IEEEtran}

\usepackage{epsfig,graphics,subfigure,psfrag,amsmath,float,subfigure}
\usepackage{latexsym,amssymb,amsmath,epsfig,subfigure}
\usepackage{cite}

{\newtheorem{Thm}{Theorem}}
\newtheorem{Lem}{Lemma}

\newtheorem{Def}{Definition}

\title{Balancing Egoism and Altruism on Interference Channel: The MIMO Case}
\author{Zuleita K.~M.~Ho and David Gesbert\\
Eurecom \\
2229 Route des Cr\^etes, 06560 Sophia Antipolis, France \\
\{hokm, gesbert\}@eurecom.fr}
\IEEEaftertitletext{\vspace{-2\baselineskip}}

\begin{document}
\maketitle
%\markboth{\MakeLowercase{\textit{submitted to}} IEEE Journal on Selected Areas In Communications}{}
\begin{abstract}
This paper considers the so-called Multiple-Input-Multiple-Output interference channel (MIMO-IC)
which has relevance in applications such as multi-cell coordination in
cellular networks as well as spectrum sharing in cognitive radio networks among
others. We address the design of precoding (i.e. beamforming) vectors at each
sender with the aim of striking a compromise
between beamforming gain at the intended receiver (Egoism) and the mitigation of
interference created towards other receivers (Altruism). Combining egoistic and altruistic beamforming has been shown previously to be instrumental to optimizing the rates in a Multiple-Input-Single-Output (MISO) interference channel (i.e. where receivers have no interference canceling capability) \cite{Jorswieck2008a, Ho2008}. Here we explore these game-theoretic concepts in the more general context of MIMO channels and use the framework of Bayesian games \cite{Myerson} which allows us to derive (semi-)distributed precoding techniques. We draw parallels with important existing work on the MIMO-IC, including rate-optimizing and interference-alignment precoding techniques, and show how such techniques may be re-interpreted through a common prism based on balancing egoistic and altruistic beamforming. Our analysis and simulations attest the improvements in terms of complexity and performance.

\end{abstract}

\section{Introduction}
The mitigation of interference in multi-point to multi-point radio systems is of utmost
importance in  contexts such as cognitive radio and multi-cell MIMO systems with full frequency reuse. 
We model a network of $N_c$
interfering radio links where each link consists of a sender trying to
communicate messages to a unique receiver in spite of the interference arising
from or created towards other links. Recently, the attention of the research
community was drawn to the so-called {\em coordinated} transmission methods where
interference effects are mitigated or even exploited in exchange
for an additional overhead in exchanging data symbols and channel state 
information (CSI) between the transmitters.

In a scenario where
the back-haul network cannot support a complete sharing of data symbols across all transmitters,  the channel is a so-called {\em
interference-channel} whereby the senders can resort to a milder form of
coordination that does not require joint encoding of data packets. 
Coordination over the interference channel may take place over one or several domains characterizing the
transmission parameters of each sender such as the choice of power levels \cite{Gesbert2008},
beamforming vectors \cite{Ho2008,Ye2003,Tenenbaum2008,Jafar2008,Heath2009},
assigned subcarriers in OFDMA \cite{Prasad2009}, scheduling \cite{Kiani2008, Choi2006} etc to cite a few.

Recently an interesting framework for beamforming-based coordination was proposed for the MISO case by which the transmitters (e.g. the base stations)
seek to strike a compromise between selfishly serving their users while ignoring the interference effects on the one
hand,  and altruistically minimizing the harm they
cause to other non-intended receivers on the other hand.  An important result in this area was the characterization of all so-called Pareto optimal beamforming solutions for the two-cell case in  the form of positive linear  combinations of the  purely selfish and purely
altruistic beamforming solutions \cite{Jorswieck2008a,Ho2008}. Unfortunately, how or whether at all this analysis can be extended to the context of MIMO-IC (i.e. where receivers have themselves multiple antennas and interference canceling capability) remains an open question. 

Coordination on the MIMO-IC has emerged as a very popular topic recently, with several important contributions shedding light on rate-scaling optimal precoding strategies based on so-called interference alignment \cite{Jafar2008,Heath2009} and rate-maximizing precoding strategies \cite{Ye2003,Wolfgang2009}, to cite just a few. In this paper, our contributions are as follows: 
\begin{itemize}
\item  We re-visit the problem of precoding on the MIMO-IC through the prism of game-theoretic egoistic and altruistic beamforming methods.
For doing so, we derive analytically the equilibria for so-called egoistic and altruistic bayesian games \cite{Myerson}.
\item We derive a game-theoretic interpretation of previous work aimed at maximizing the sum-rate over the MIMO-IC, such as \cite{Wolfgang2009}.
\item We propose a new simplified precoding technique aimed at sum rate maximization, based on balancing the egoistic and the altruistic behavior at each transmitter, where the balancing weights are derived from statistical parameters.
\item We show that our algorithm exhibits the same optimal rate scaling (when SNR grows) as shown by recent interesting iterative interference-alignment based methods \cite{Jafar2008,Heath2009}. At finite SNR, we show improvements in terms of sum rate, especially in the case of asymmetric networks where interference-alignment methods are unable to properly weigh the contributions on the different interfering links to the sum rate.   
\end{itemize}
%%%%%%%%%%%

\subsection{Notations}
The lower case bold face letter represents a vector whereas the upper case bold face letter represents a matrix. $(.)^H$ represents the complex conjugate transpose. $\mathbf{I}$ is the identity matrix. $V^{(max)}(\mathbf{A})$ (resp. $V^{(min)}(\mathbf{\mathbf{A}})$) is the eigenvector corresponding to the largest (resp. smallest) eigenvalue of $\mathbf{\mathbf{A}}$. $\mathcal{E}_{B}$ is the expectation operator over the statistics of the random variable $B$. $\mathbb{S} \setminus \mathbb{B}$ define a set of elements in $\mathbb{S}$ excluding the elements in $\mathbb{B}$. 

 \section{System Model}
We study a wireless network of $N$ cells, where a subset of $N_c \leq N$ transmitters will form a coordination cluster (i.e. will be coordinated across) and are especially considered. The transmitters could be the base stations (BS) in the cellular downlink. Each transmitter
is equipped with $N_t$ antennas and the receivers (e.g. mobile stations)  
 with $N_r$ antennas. In each of the $N_c$ cells, an orthogonal multiple access scheme is assumed, hence each transmitter (Tx) communicates with a unique receiver (Rx) at a time.  Transmitters are not allowed or able to exchange user message information, giving rise to an interference channel over which we seek some form of beamforming-based coordination.  The channel from Tx $i$ to Rx $j$
 $\mathbf{H}_{ji} \in \mathcal{C}^{N_r \times N_t}$ is given by:
 \begin{equation}\label{eqt:channel}
 	\mathbf{H}_{ji}= \sqrt{\alpha_{ji}} \bar{\mathbf{H}}_{ji}
 \end{equation}
 
Each element in channel matrix $\bar{\mathbf{H}}_{ji}$ is an
independent identically distributed complex Gaussian random variable with zero
mean and unit variance and $\alpha_{ji}$ denotes the slow-varying shadowing and pathloss attenuation.

The transmit beamforming vector of Tx $i$ is $\mathbf{w}_i \in \mathcal{C}^{N_t
\times 1}$ and the receive beamforming vector of Rx $i$ is $\mathbf{v}_i \in
\mathcal{C}^{N_r \times 1}$. As in several important contributions dealing with coordination on the interference channel, we assume linear precoding (beamforming)
\cite{Ho2008, Choi2006, Jorswieck2008a,  Jafar2008}.  With the noise
variance $\sigma_{i}^2$ at Rx $i$ and transmit power $P$, the received signal-to-noise ratio of Rx $i$ is
\begin{equation}\label{eqt:sinr}
	\gamma_i= \frac{|\mathbf{v}_i^H \mathbf{H}_{ii} \mathbf{w}_i|^2 P}{
\sum_{j \neq i}^{N_c} |\mathbf{v}_i^H
\mathbf{H}_{ij} \mathbf{w}_j|^2 P + \sigma_{i}^2}.
\end{equation}
\subsection{Receiver design}
The receivers are assumed to employ maximum SINR (Max-SINR) beamforming throughout the paper so as to also maximize their rates \cite{Paulraj2003}. The receive beamformer is classically given by:
\begin{equation}\label{eqt:max_sinr}
\mathbf{v}_i= \frac{{C_{Ri}}^{-1} \mathbf{H}_{ii}
\mathbf{w}_i}{|{C_{Ri}}^{-1} \mathbf{H}_{ii} \mathbf{w}_i|}
\end{equation} where $C_{Ri}$ is the covariance matrix of received interference and
noise at Rx $i$ and $C_{Ri}= \sum_{j \neq i} \mathbf{H}_{ij} \mathbf{w}_j \mathbf{w}_j^H
\mathbf{H}_{ij}^H P+ \sigma_{i}^2 \mathbf{I}$.

Importantly, the noise will in practice capture thermal noise effects but also any interference originating from the rest of the network, i.e. coming from transmitters located beyond the coordination cluster. Thus, depending on path loss and shadowing effects, the $ \{\sigma_{i}^2 \}$  may be quite different from each other \cite{molisch}.

\subsection{Limited Channel knowledge}	
\begin{figure}
\begin{center}
 	\includegraphics[width=8cm, height=6cm]{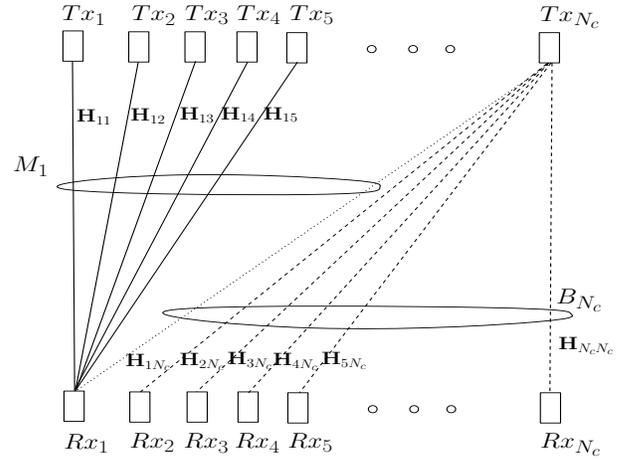}
 	\caption{Limited channel knowledge model for  an example of transmitter, here TX $N_c$, indicated by dotted lines, and an example of receiver, here RX 1, indicated by solid lines.}\label{fig:channel_model_bs}
\end{center}
 \end{figure}

To allow for overhead reduction and a better scalability of multi-cell coordination techniques when the number of coordinated links $N_c$ is large, we seek solutions which can operate based on limited, preferably local, CSI. Although there may exist various ranges and definitions of local CSI, we assume the devices (Tx and Rx alike) are able to  gain direct knowledge of those channel coefficients directly connected to them, as illustrated in Fig. \ref{fig:channel_model_bs}. 

The set of CSI locally available (resp. not available) at Tx $i$ by $\mathbb{B}_i$ (resp. $\mathbb{B}_i^\perp$) is defined by:
$\mathbb{B}_i= \left\{ \mathbf{H}_{ji} \right\}_{j=1, \ldots, N_c} \; ;
\; \mathbb{B}_i^\perp= \left\{ \mathbf{H}_{kl}\right\}_{k,l=1 \ldots N_c}
\setminus \mathbb{B}_i. $
Similarly,  define the set of channels known (resp. unknown) at Rx $i$ by $\mathbb{M}_i$ (resp. $\mathbb{M}_i^\perp$) by:
$\mathbb{M}_i= \left\{ \mathbf{H}_{ij} \right\}_{j= 1, \ldots, N_c} \; ;
\; \mathbb{M}_i^\perp = \left\{ \mathbf{H}_{kl}\right\}_{k,l=1 \ldots N_c}
\setminus \mathbb{M}_i.$

\emph{Additional receiver feedback}:
Because local CSI is insufficient to exploit all the degrees of freedom of the MIMO-IC \cite{Jafar2008}, some additional limited feedback will be considered where indicated, in the form of  feedback of the beamforming vectors ${\bf v}_i$ used at the receiver. Of course, in the case of reciprocal channels, the feedback requirement can be replaced by a channel estimation step based on uplink pilot sequences. Additionally, it will be classically assumed that the receivers are able to estimate the covariance matrix of their interference signal  based on  transmitted pilot sequences.

 \section{Bayesian Games on interference channel}\label{section:bayesian}
Bayesian games are a class of games in which players must optimize their strategy based on
\emph{incomplete state information} \cite{Myerson} and hence are particularly well suited to distributed optimization problems. Below we provide a few useful definition for this framework in the context
of the MIMO-IC.

 A Bayesian game is defined as the following, 
 \begin{equation}
 	G= < \mathcal{N},\Omega, <\mathcal{A}_i, u_i, \mathbb{B}_i^\perp>_{i
\in N}>
 \end{equation}
where $\mathcal{N}$ is the set of players in the game, here refers to the set of transmitters $\left\{ 1, \ldots, N_c \right\}$ .
$\Omega$ is the set of all possible global channel states $\left\{\mathbb{C}^{N_r \times N_t }\right\}^{N_c}$. $\mathcal{A}_i$ is the action set of player $i$, here refers to all choice of beamforming vectors $\mathbf{w}_i$ such that the power constraint is fulfilled $|\mathbf{w}_i|^2\leq 1$. $u_i: \Omega \times \mathcal{A}_i \rightarrow \mathbb{R}$ is the
utility function of player $i$. In the next section we define egoistic and altruistic utilities. $\mathbb{B}_i^\perp$ is the \emph{missing} channel state information  at player $i$.
\begin{Def}
	A strategy of player $i$, here refers to beamforming design, $\mathbf{w}_i: \mathbb{B}_i \rightarrow \mathcal{A}_i$ is a
deterministic choice of action given information $\mathbb{B}_i$ of player $i$. 
\end{Def}
\begin{Def}
	A strategy profile $\mathbf{W}^*=(\mathbf{w}_i^*, \mathbf{w}_{-i}^*)$
achieves the Bayesian Equilibrium if $\mathbf{w}_i^*$ is the best response of player $i$, here optimal transmit beamformer of player $i$, given
strategies $\mathbf{w}_{-i}^*$ for all other players and is characterized by
\begin{equation}
	\forall i \; \;	\mathbf{w}_i^*=\arg \max
\mathcal{E}_{\mathbb{B}_i^\perp} \left\{ u_i(\mathbf{w}_i, \mathbf{w}_{-i}^*, \mathbb{B}_i, \mathbb{B}_i^\perp) \right\} 
\end{equation}
\end{Def}
Note that, intuitively, the player's strategy is optimized by averaging over the distribution of all missing channel state information.

In the following sections, we derive the equilibria for so-called {\em egoistic}  and {\em altruistic} bayesian games respectively. These equilibria contribute to extreme strategies which do not perform optimally in terms of the overall network performance, yet can be exploited as components of more general  beamforming-based coordination techniques. 
 
 \section{Bayesian Games with Receiver Beamformer Feedback}
We assume that each Tx has the local channel state information and the added
knowledge of receive beamformers through a feedback channel. Under this assumption, we  analyze the Egoistic and
Altruistic beamforming solutions.

\subsection{Egoistic Bayesian Game}
Given receive beamformers as a common knowledge, the best response strategy of Tx $i$ which maximizes the utility function, its own SINR, 
\begin{equation}\label{def:ego_fb}
 u_i(\mathbf{w}_i,
\mathbf{w}_{-i}, \mathbb{B}_i, \mathbb{B}_i^\perp)= \frac{ |\mathbf{v}_i^H \mathbf{H}_{ii}
\mathbf{w}_i|^2 P}{\sum_{j\neq i}^{N_c} |\mathbf{v}_i^H \mathbf{H}_{ij} \mathbf{w}_j|^2 P
+\sigma_i^2}
\end{equation} is the following:

\begin{Thm}\label{thm:ego_fb}
 The best-response strategy of Tx $i$ in the egoistic Bayesian game is 
\begin{equation}
	\mathbf{w}_i^{Ego}= V^{(max)}(\mathbf{E}_i)
\end{equation} where $\mathbf{E}_i$ will denote the {\em egoistic equilibrium matrix} for Tx $i$, given by 
$\mathbf{E}_i=\mathbf{H}_{ii}^H \mathbf{v}_i \mathbf{v}_i^H \mathbf{H}_{ii}$
 and the corresponding receiver is given by $\mathbf{v}_i= \frac{{C_{Ri}}^{-1} \mathbf{H}_{ii} \mathbf{w}_i^{Ego}}{|{C_{Ri}}^{-1} \mathbf{H}_{ii} \mathbf{w}_i^{Ego}|}$

\end{Thm}
\begin{proof}
The knowledge of receive beamformers decorrelates the maximization problem. The
maximization problem can be written as $\mathbf{w}_i^{Ego}=\arg \max_{|\mathbf{w}_i|
\leq 1} \mathcal{E}_{\mathcal{B}_i^\perp} \left\{\frac{1}{\sum_{j\neq i}^{N_c}
 |\mathbf{v}_i^H \mathbf{H}_{ij} \mathbf{w}_j|^2 P +\sigma_i^2} \right\} \mathbf{w}_i^H
\mathbf{E}_i \mathbf{w}_i$.
The egoistic-optimal transmit beamformer is  the dominant eigenvector  $\mathbf{w}_i^{Ego}=V^{(max)}(\mathbf{E}_i)$.
\end{proof}

 \subsection{Altruistic Bayesian Game}
The  altruistic utility at Tx $i$ is defined here in the sense of  minimizing the expectation of the sum of
interference power towards other Rx's. 
\begin{equation}\label{def:alt}
	u_i(\mathbf{w}_i, \mathbf{w}_{-i}, \mathbb{B}_i, \mathbb{B}_i^\perp)= -\sum_{j \neq
i}|\mathbf{v}_j^H \mathbf{H}_{ji} \mathbf{w}_i|^2	
\end{equation}

\begin{Thm}
	 The
best-response strategy of Tx $i$ in the altruistic Bayesian game is given by:
\begin{equation}
	\mathbf{w}_i^{Alt}= V^{(min)}(\sum_{j \neq i} \mathbf{A}_{ji})
\label{eqt:mmse}	
\end{equation}
where $\mathbf{A}_{ji}$ will denote the {\em altruistic equilibrium matrix for Tx $i$ towards Rx $j$}, defined by $\mathbf{A}_{ji}=\mathbf{H}_{ji}^H \mathbf{v}_j \mathbf{v}_j^{H} \mathbf{H}_{ji}$. The corresponding receiver is $\mathbf{v}_i= \frac{C_{Ri}^{-1} \mathbf{H}_{ii} \mathbf{w}_i}{|C_{Ri}^{-1} \mathbf{H}_{ii} \mathbf{w}_i|} $.
\end{Thm}

\begin{proof}
 The altruistic utility can be rewritten as -$\sum_{j \neq i} |\mathbf{v}_j^H \mathbf{H}_{ji} \mathbf{w}_i|^2= -\sum_{j \neq i} \mathbf{w}_i^H \mathbf{A}_{ji} \mathbf{w}_i$. Since the $\mathbf{v}_j$ are known from feedback, the optimal $\mathbf{w}_i$ is the least dominant eigenvector of the matrix $\sum_{j \neq i}\mathbf{A}_{ji}$.
\end{proof}

 \section{Sum Rate Maximization with Receive Beamformer Feedback}

From the results above, it can be seen that balancing altruism and egoism for player $i$ can be done by trading-off between the dominant eigenvectors of the egoistic equilibrium $\mathbf{E}_i$ and negative altruistic equilibrium  $\left\{ \mathbf{A}_{ji} \right\}$ ($j \neq i$) matrices.
Interestingly, it can be shown that sum rate maximizing precoding for the MIMO-IC does exactly that. Thus we hereby briefly re-visit rate-maximization approaches such as \cite{Wolfgang2009} with this perspective. 
  
Denote the sum rate by $\bar{R}=\sum_{ i=1}^{N_c} R_i$ where $	R_i = \log_2 \left(1+ \frac{|\mathbf{v}_i^H \mathbf{H}_{ii}
\mathbf{w}_i|^2 P }{\sum_{j \neq i}^{N_c} |\mathbf{v}_i^H
\mathbf{H}_{ij} \mathbf{w}_j|^2 P + \sigma_i^2} \right)$.

\begin{Lem}\label{Lem:opt_lambda}
	The transmit beamforming vector which maximizes the sum rate $\bar{R}$ is given by the following dominant eigenvector problem, 
\begin{equation}\label{eqt:ego_alt_linear}
	\left(\mathbf{E}_i + \sum_{j \neq i}^{N_c} \lambda_{ji}^{opt} \mathbf{A}_{ji}
\right) \mathbf{w}_i= \mu_{max} \mathbf{w}_i
\end{equation} where real values $\lambda_{ji}^{opt}$, $\mu_{max}$ are defined in the proof.
\end{Lem}

\begin{proof}
see appendix \ref{app:opt_lam}
\end{proof}
Note that the balancing between altruism and egoism in sum rate maximization is done using a simple {\em linear combination} of the altruistic and egoistic equilibrium matrices.
The balancing parameters, $ \{\lambda_{ji}^{opt} \}$, coincide with the pricing parameters invoked in the iterative algorithm proposed in \cite{Wolfgang2009}.
Clearly, these parameters plays a key role, however their computation  is a function of the {\em global} channel state information.
 Instead we seek below a suboptimal egoism-altruism balancing technique which only requires statistical channel information, while exhibiting the right performance scaling.  
 
\subsection{Egoism-altruism balancing algorithm: DBA-RF}
We are proposing the following distributed beamforming algorithm with receiver feedback (DBA-RF),  to compute the transmit beamformers
\begin{equation}\label{eqt:w_lb}
\mathbf{w}_i = V^{max} \left(\mathbf{E}_i + \sum_{j \neq i}^{N_c}  \lambda_{ji} \mathbf{A}_{ji} \right) .
\end{equation}

\emph{DBA-RF} iterates between transmit and receive beamformers in a way similar to recent interference-alignment based methods such as e.g. \cite{Jafar2008,Heath2009}. However here, interference alignment is {\em not} a design criterion. In  \cite{Jafar2008}, an improved interference alignment technique based on alternately maximizing the SINR at both sides is proposed. In contrast, the Max-SINR criterion is only used at the receiver side. This distinction is important as it dramatically changes performance in certain situations (see Section \ref{section:result}).

One important aspect of the algorithm above is whether it fully exploits the degree of freedom of the interference channel as shown per \cite{Jafar2008}, i.e. whether it achieves the so-called interference alignment in high SNR regime. The following theorem answers this question positively.
\begin{Def}
Interference is aligned when the following equations are satisfied at the same time \cite{Jafar2008}:
\begin{equation}\label{def:IA}
\mathbf{v}^H_i \mathbf{H}_{ij} \mathbf{w}_j =0 \; \; \forall i,j \neq i 
\end{equation}
\end{Def}

\begin{Def}
Define the set of beamforming vectors solutions in downlink (respectively uplink) interference alignment to be \cite{Jafar2008}
\begin{small}
\begin{eqnarray}
& &\mathcal{IA}^{DL} = \label{def:low_rank}\\  
\nonumber & & \;\left\{ (\mathbf{w}_1, \ldots, \mathbf{w}_{N_c}): \sum_{k \neq i}^{N_c} \mathbf{H}_{ik} \mathbf{w}_k \mathbf{w}_k^{H} \mathbf{H}_{ik}^H \text{ is low rank, } \forall i \right\}\\
\nonumber & &\mathcal{IA}^{UL} = \\
\nonumber & & \;\left\{ (\mathbf{v}_1, \ldots, \mathbf{v}_{N_c}): \sum_{k \neq i}^{N_c} \mathbf{H}_{ki}^H \mathbf{v}_k \mathbf{v}_k^{H} \mathbf{H}_{ki} \text{ is low rank, } \forall i \right\}.
\end{eqnarray}
\end{small}
Thus, for all $(\mathbf{w}_i,\ldots, \mathbf{w}_{N_c}) \in \mathcal{IA}^{DL}$, there exist receive beamformers $\mathbf{v}_i, i=1,\ldots,N_c$ such that \eqref{def:IA} is satisfied.
\end{Def}
Note that the uplink alignment solutions are defined for a virtual uplink having the same frequency and only appear here as technical concept helping with the proof. 

\begin{Thm}
Assume the downlink interference alignment set is non empty (IA is feasible). Denote  average SNR of link $i$ by $\gamma_i=\frac{P \alpha_{ii}}{\sigma_i^2}$. Let $ \lambda_{ji} = -\frac{1+\gamma_i^{-1}}{1+\gamma_j^{-1}}\gamma_j$, then in the large SNR regime, $P \rightarrow \infty$ , any transmit beamforming vector in $\mathcal{IA}^{DL}$ is a convergence (stable) point of \emph{DBA-RF}.
\end{Thm}
\begin{proof}
We provide here a sketch of the proof. For full details, please refer to \cite{Ho2009}. To prove that \emph{IA} is a convergence point of \emph{DBA-RF}, we would prove that once \emph{DBA-RF} achieves interference alignment, \emph{DBA-RF} will not deviate from the solution.

Assumed interference alignment is reached and let $(\mathbf{w}_1^{IA}, \ldots, \mathbf{w}_{N_c}^{IA}) \in \mathcal{IA}^{DL}$ and $(\mathbf{v}_1^{IA}, \ldots, \mathbf{v}_{N_c}^{IA}) \in \mathcal{IA}^{UL}$. Let $\mathbf{Q}_i^{DL}=\sum_{k \neq i}^{N_c} \mathbf{H}_{ik} \mathbf{w}_k^{IA} \mathbf{w}_k^{IA,H} \mathbf{H}_{ik}^H$ and $\mathbf{Q}_i^{UL}=\sum_{k \neq i}^{N_c} \mathbf{H}_{ki}^H \mathbf{v}_k^{IA} \mathbf{v}_k^{IA,H} \mathbf{H}_{ki}$.

At the transmitters: In high SNR regime, $\lambda_{ji}$ becomes negative infinity and \emph{DBA-RF} gives $\mathbf{w}_i = V^{min}( \mathbf{Q}_i^{UL})$ \eqref{eqt:w_lb}. By  \eqref{def:low_rank}, $\mathbf{Q}_i^{UL}$ is low rank and thus $\mathbf{w}_i$ is in the null space of $\mathbf{Q}_i^{UL}$. In direct consequences, the conditions of \emph{IA} \eqref{def:IA} are satisfied. Thus, $(\mathbf{w}_1, \ldots, \mathbf{w}_{N_c}) \in \mathcal{IA}^{DL}$.

At the receivers:  The receive beamformer is defined as $\mathbf{v}_i=\arg \max \frac{\mathbf{v}_i^H \mathbf{H}_{ii} \mathbf{w}_i \mathbf{w}_i^H \mathbf{H}_{ii}^H \mathbf{v}_i}{\mathbf{v}_i^H \mathbf{Q}_i^{DL} \mathbf{v}_i}$. Since $\mathbf{Q}_i^{DL}$ is low rank, the optimal $\mathbf{v}_i$ would make the denominator zero and thus, $\mathbf{v}_i$ is in the null space of $\mathbf{Q}_i^{DL}$. Hence, $\mathbf{v}_i \in \mathcal{IA}^{UL}$.

Since both $\mathbf{w}_i$ and $\mathbf{v}_i$ stays within $\mathcal{IA}^{DL}$ and $\mathcal{IA}^{UL}$, \emph{IA} is a convergence point of DBA-RF in high SNR.

\end{proof}

\section{Simulation Results}\label{section:result}
In this section, we investigate the sum rate performance of \emph{DBA-RF} in comparison with several related methods, namely the \emph{Max-SINR} method \cite{Jafar2008}, the alternated-minimization  (\emph{Alt-Min}) method for interference alignement \cite{Heath2009} and the sum rate optimization method (\emph{SR-Max}) \cite{Wolfgang2009}.  The \emph{SR-Max} method is by construction optimal but is more complex and requires extra sharing or feedback of pricing information among the transmitters.  Also, \emph{Max-SINR} method in \cite{Jafar2008} does not aim to null out interference, but maximize receive SINR instead.

User located in the cells follow a uniform distribution. To ensure a fair comparison, all the algorithms in comparisons are initialized to the same  solution and have the same stopping condition. We perform sum rate comparisons in both symmetric channels and asymmetric channels where links undergo different levels of out-of-cluster noise.  Define the Signal to Interference ratio of link $i$ to be $SIR_i= \frac{\alpha_{ii}}{ \sum_{j \neq i}^{N_c} \alpha_{ij}}$. The $SIR$ is assumed to be 10dB for all links, unless otherwise stated. Denote the difference in SNR between two links in asymmetric channels by $\Delta SNR$. 

\subsection{Symmetric Channels}

\begin{figure}
  \begin{center}
   \includegraphics[height=7cm, width=9cm]{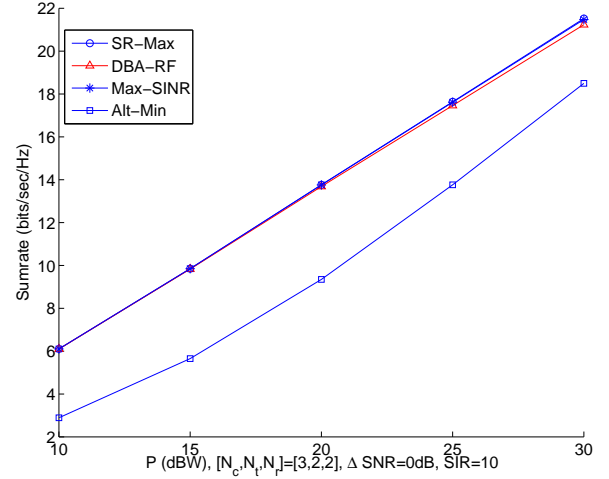}
 \caption{Sum rate comparison in multi links systems with $[N_c,N_t,N_r]=[3,2,2]$ with increasing SNR. DBA-RF achieves close to optimal performance.}\label{fig:sumrate_lam_nc3nt2nr2}
  \end{center}
 \end{figure}

Fig. \ref{fig:sumrate_lam_nc3nt2nr2} illustrates the sum rate comparison of \emph{DBA-RF} with \emph{Max-SINR}, \emph{Alt-Min} and \emph{SR-Max} in a system of 3 links and each Tx and Rx have 2 antennas. Since interference alignment is feasible in this case, the sum rate performance of \emph{SR-Max} and \emph{Max-SINR} increase linearly with SNR. \emph{DBA-RF} achieves sum rate performance with the same scaling as \emph{Max-SINR} and \emph{SR-Max}. 

\begin{figure}
  \begin{center}
   \includegraphics[height=7cm, width=9cm]{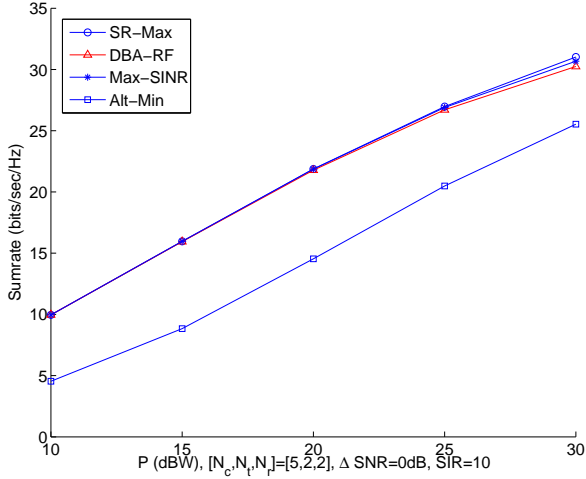}
 \caption{Sum rate comparison in multi links systems with $[N_c,N_t,N_r]=[5,2,2]$ with increasing SNR. \emph{DBA-RF}  achieves close to optimal performance.} \label{fig:sumrate_nc5nt2nr2}
  \end{center}
 \end{figure}

 In Fig. \ref{fig:sumrate_nc5nt2nr2}, we show the sum rate in a system of 5 links where each Tx and Rx are equipped wtih 2 antennas. Note that in this case interference alignment is \emph{infeasible}. The sum rate performances saturate at high SNR regime. \emph{DBA-RF} achieves close to optimal performance in spite of the unfeasibility of interference alignment.
 
 \subsection{Asymmetric Channels}
\begin{figure}
  \begin{center}
   \includegraphics[height=7cm, width=9cm]{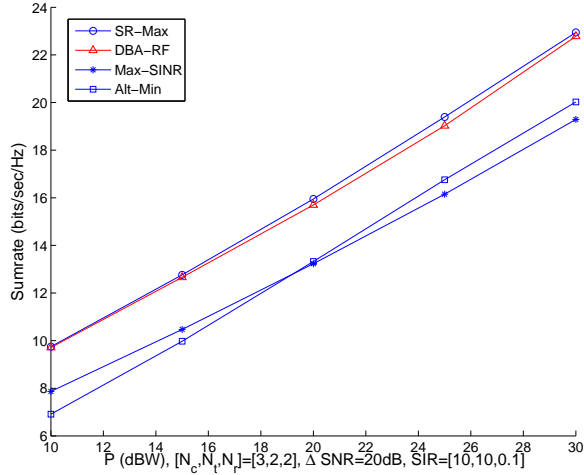}
 \caption{Sum rate performance for asymmetric channel of 3 links system. \emph{DBA-RF} outperforms most algorithms by balancing Egoism and Altruism.}\label{fig:asymmetric}
  \end{center}
 \end{figure}

In the asymmetric system, some links undergo uneven levels of noise. In Fig. \ref{fig:asymmetric}, we compare the sum rate performance in a 3 links system where SNR of link 1 and link 2 is larger than that of link 3 by $\Delta SNR=20dB$. The SIR's of the links are $[10,10,0.1]$ respectively. Thus  link 3  not only suffers from strong noise, but also a strong interference channel. The asymmetry penalizes the Max-SINR and interference alignment methods because they are unable to properly weigh the contributions of the weaker link in the sum rate. The Max-SINR strategy turns out to make link 3 very egoistic in this example, while its proper behavior should be altruistic.
In contrast, \emph{DBA-RF} exploits useful statistical information, allowing weaker link to allocate their spatial degrees of freedom wisely toward stronger links  and vice versa, toward a better sum rate.

 \section{Conclusion}
 We derive the equilibria for the egoistic and altruistic bayesian games. We suggest a precoding technique based on balancing the egoistic and the altruistic behavior at each transmitter with the aim of maximizing the sum rate.  We obtain an iterative beamforming algorithm which exhibits the same optimal rate scaling (when SNR grows) shown by recent iterative interference-alignment based methods. By simultaneously equilibrating egoistic and altruistic solutions for all links, we are able to obtain close to optimum performance in situations with both symmetric and asymmetric link quality levels.
\section{Appendix}

\subsection{Proof of Lemma \ref{Lem:opt_lambda}}
\label{app:opt_lam}
 Define the largrangian of the sum rate maximization problem to be 
$\mathcal{L}(\mathbf{w}_i ,\mu)= \bar{R} - \mu_{max}( \mathbf{w}_i^H \mathbf{w}_i -1)$.
The neccessary condition of largrangian $\frac{\partial}{\partial
\mathbf{w}_i^H}\mathcal{L}(\mathbf{w}_i ,\mu)=0$ gives:
$\frac{\partial}{\partial \mathbf{w}_i^H} R_i + \sum_{j \neq i}^{N_c}
\frac{\partial}{\partial \mathbf{w}_i^H} R_j = \mu_{max} \mathbf{w}_i 
$. With elementary matrix calculus, 
$\frac{\partial}{\partial \mathbf{w}_i^H} R_i  =
\frac{P}{\sum_{j=1}^{N_c} |\mathbf{v}_i^H \mathbf{H}_{ij}
\mathbf{w}_j|^2 P + \sigma_i^2} \mathbf{E}_i \mathbf{w}_i$ and 
$\frac{\partial}{\partial \mathbf{w}_i^H} R_j = -\frac{|\mathbf{v}_j^H \mathbf{H}_{jj} \mathbf{w}_j|^2 P}{\sum_{k=1}^{N_c} |\mathbf{v}_j^H \mathbf{H}_{jk}
\mathbf{w}_k|^2 P  + \sigma_j^2} \frac{P }{\sum_{k \neq
j}^{N_c} |\mathbf{v}_j^H \mathbf{H}_{jk} \mathbf{w}_k|^2 P +
\sigma_j^2} \mathbf{A}_j \mathbf{w}_i.$

%where $P$ is the transmit power and $\alpha_{ij}$ is the path
%loss from BS $j$ to MS $i$ defined in \eqref{eqt:channel}

 Thus, $\lambda_{ji}^{opt}$ is a function of all channel states information and
beamformer feedback, $\lambda_{ji}^{opt} = -\frac{|\mathbf{v}_j^H \mathbf{H}_{jj}
\mathbf{w}_j|^2 P}{\sum_{k=1}^{N_c} |\mathbf{v}_j^H \mathbf{H}_{jk}
\mathbf{w}_k|^2 P  + \sigma_j^2} \frac{\sum_{j=1}^{N_c}
|\mathbf{v}_i^H \mathbf{H}_{ij}
\mathbf{w}_j|^2 P  + \sigma_i^2}{\sum_{k \neq
j}^{N_c} |\mathbf{v}_j^H \mathbf{H}_{jk} \mathbf{w}_k|^2 P +
\sigma_j^2}.$

\bibliography{zuleitabib1}{} 
\bibliographystyle{IEEEbib}

\end{document}